\RequirePackage{lineno}

\documentclass[3p,times]{elsarticle}

\usepackage{ecrc}

\volume{00}

\firstpage{1}

\journalname{Nuclear Physics A}

\runauth{}

\jid{nupha}

\usepackage{amssymb}

\usepackage[figuresright]{rotating}

\newcommand{\pT} {\ensuremath{p_{\mathrm{T}}}}

\begin{document}

\begin{frontmatter}



\dochead{}

\title{Measurement of $v_{n}$ coefficients and event plane correlations in 2.76 TeV Pb-Pb collisions with ATLAS}


\author{Soumya Mohapatra\footnote{$Email$ $address$: soumya@cern.ch} on behalf of the ATLAS Collaboration}

\address{Departments of Physics $\&$ Astronomy and Chemistry, Stony Brook University, NY 11794, USA}

\begin{abstract}
Measurements of flow harmonics $v_1-v_6$ are presented in a broad range of transverse momentum and centrality for $\sqrt{s_{NN}} = 2.76$ TeV Pb-Pb collisions with the ATLAS experiment. The fourier coefficients of two-particle correlations are shown to factorize as products of the single particle $v_n$ for $n=2-6$. The factorization breaks for $n=1$ due to effects of global momentum conservation. The dipolar $v_1$ associated with the initial dipole asymmetry is extracted from the two-particle correlations via a two component fit that accounts for momentum conservation. The magnitude of the extracted dipolar $v_1$ is comparable to $v_3$. Measurements of correlations between harmonic planes of different orders, which give additional constraints on the initial geometry and expansion mechanism of the medium produced in the heavy ion collisions, are also presented.
\end{abstract}


\end{frontmatter}


\section{Introduction}
In heavy-ion collisions the produced fireball has spatial anisotropies of many orders: elliptic triangular etc. These anisotropies are transferred from position space to momentum space due to the collective expansion of the medium or path-length dependent suppression of particles. This results in the azimuthal distribution of the final particle yields being modulated by the flow harmonics $v_{n}$ about the different harmonic planes $\Phi_{n}$ \cite{reaction_plane_defns}:

\begin{equation}
\frac{dN}{d\phi}\propto1+2\sum_{n=1}^{\infty}v_{n}(\pT,\eta)\cos(n\phi-n\Phi_{n})\label{eq_flow_equation}
\end{equation}

The $v_n$ can be measured by the event plane (EP) method by correlating the single-particle azimuthal distributions with the harmonic planes $\Phi_n$. They can also be measured by the two-particle correlation (2PC) method where the particle-pair distribution in relative azimuthal angle $\Delta\phi=\phi_{a}-\phi_{b}$ are measured (the subscripts $a$ and $b$ label the two particles, commonly called trigger particle and partner particle).  The 2PC correlation function can be expanded in a Fourier series in $\Delta\phi$ as:

\begin{equation}
\frac{dN_{\mathrm{pairs}}}{d\Delta\phi}\propto1+2\sum_{n=1}^{\infty}v_{n,n}\cos(n\Delta\phi)\label{eq_correlation}
\end{equation}

\noindent  If the correlations are dominated by single particle anisotropies (Eq. \ref{eq_flow_equation}), then the Fourier coefficients $v_{n,n}$ are equal to the product of the individual single particle $v_n$:

\begin{equation}
v_{n,n}(\pT^\mathrm{a},\pT^\mathrm{b})=v_{n}(\pT^\mathrm{a})v_{n}(\pT^\mathrm{b})\label{eq_scaling_relation}
\end{equation}

\noindent Using Eq. \ref{eq_scaling_relation}, one can obtain the $v_n$ from the 2PC. The above relation is violated if non-flow effects, such as jets and resonance decays are large. Thus the non-flow effects must be suppressed before extracting the flow harmonics from the correlations.

The flow harmonics are important observables as they contain information about the initial geometry and transport properties of the medium \cite{bib_visc_calc1,bib_visc_calc2}. A better understanding of the $v_n$ can also explain the origin of the ridge, an elongated structure along $\Delta\eta$ at $\Delta\phi\sim 0$ \cite{bib_ridge_cone1} and the so-called ``mach-cone", a double-hump structure on the away-side \cite{bib_ridge_cone2} seen in the 2PC. These were initially interpreted as the response of the medium to the energy deposited by the quenched jets. However, recent studies \cite{bib_v3cone_} have shown that higher-order flow harmonics can contribute to these structures.  

Another set of observables closely related to the $v_n$ are the correlations between the event plane angles $\Phi_n$. These correlations can be produced due to correlations between eccentricities of different orders in the initial geometry or they can develop during the dynamical expansion of the produced matter. Thus these measurements provide additional constraints on initial geometry and the expansion mechanism of the medium \cite{our_ep_paper,our_ep_paper2}.

The $v_n$ results presented are for charged hadrons reconstructed in the ATLAS inner detector \cite{atlas_detector} covering  $|\eta|<2.5$. The $\Phi_n$ planes for the $v_2$-$v_6$ measurements via the event plane method were determined using the forward calorimeter covering $|\eta|\in(3.3,4.9)$. For the event plane correlation analysis, the entire EM calorimetry covering $\eta\in(-4.9,4.9)$ was used. All measurements were done using 8 $\mu b^{-1}$ of Minimum Bias Pb-Pb data at $\sqrt{s_{NN}}$ of 2.76 TeV. Details of the results presented here are published in \cite{bib_vn_note} and \cite{bib_EP_note}.

\begin{figure}
\includegraphics[width=0.8\paperwidth]{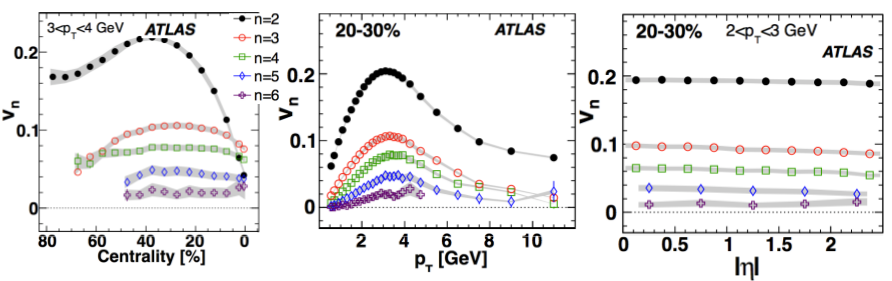}
\caption{A sample of $v_2$-$v_6$ values measured by the event plane method. The first panel shows the centrality dependence for the $v_n$. The second panel shows the $\pT$ dependence for (20-30)\% centrality and the last panel shows the $\eta$ dependence. \label{fig_ep_summary}}
\end{figure}

Figure  \ref{fig_ep_summary} summarizes the results for $v_2$ to $v_6$ measured by the event plane method. Qualitatively the following features are seen: the $v_n$ coefficients rise and fall with centrality with $v_2$ having considerable centrality dependence (being driven by the average geometry). Typically $v_2$ is much larger than the other harmonics, but in most central collisions $v_3$, $v_4$ and even $v_5$ can become larger than $v_2$. The $v_n$ coefficients rise and fall with $\pT$, this is interpreted as the driving mechanism behind the $v_n$ changing from collective expansion at low $\pT$ to path-length dependent suppression at high $\pT$. The $v_n$ coefficients are approximately boost invariant showing that they represent non-local correlations.

\section{Two-Particle Correlations}

\begin{figure}
\centering
\includegraphics[width=0.8\paperwidth]{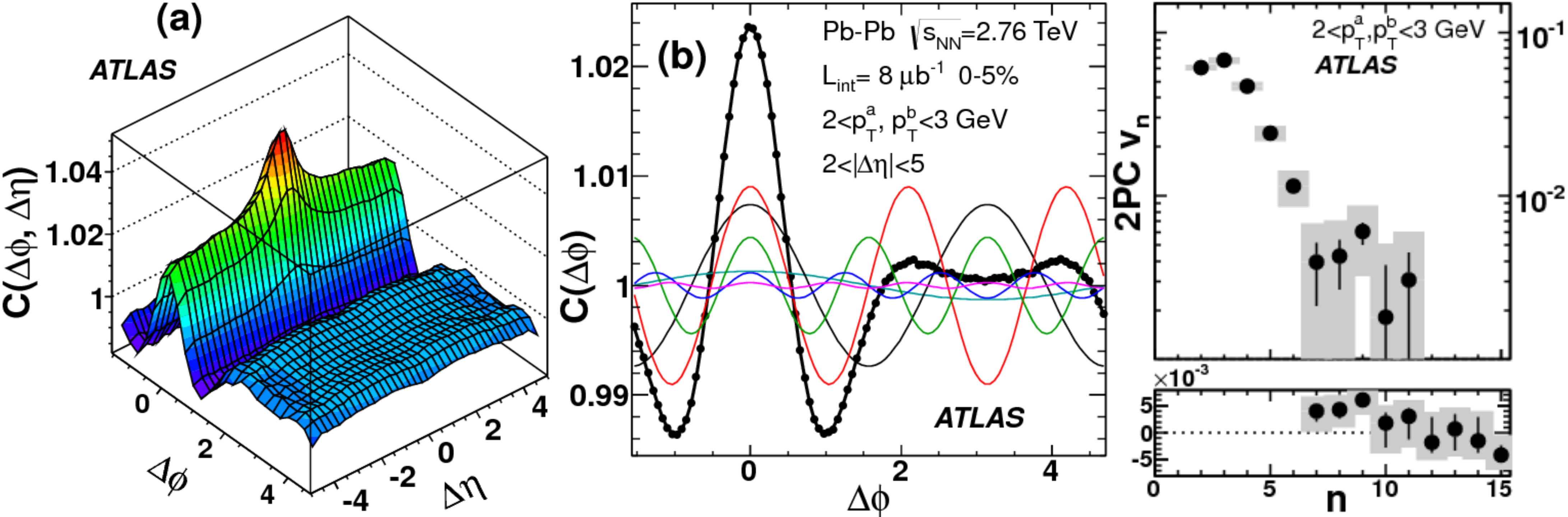}
\caption{The steps involved in the extraction of the $v_n$ for 2-3 GeV fixed-$\pT$ correlation. Panel a): the two-dimentional correlation function (shown for $|\Delta\eta|<4.75$ to reduce fluctuations near the edge). Panel b): The one-dimentional $\Delta\phi$ correlation function for $2<|\Delta\eta|<5$ (black points) overlaid with its individual Fourier components and their sum (thick black line). Panel c): $v_{n}$ vs. $n$ (The bars and bands indicate statistical and systematic uncertainties respectively). The lower sub panel in c) shows $v_n$ for $n\geq7$ but in a linear scale.}
\label{fig_2pc_method}
\end{figure}

Figure  \ref{fig_2pc_method} shows the procedure used for obtaining the $v_{n}$ from the 2PC. Panel a) shows the two-dimensional correlation function in $\Delta\eta-\Delta\phi$ for $\pT^\mathrm{a},\pT^\mathrm{b}\in(2,3)$ GeV. Such correlations, where both trigger and partner particles are in the same $\pT$ range, are termed as fixed-$\pT$ correlations. Long range structures, namely the ridge at $\Delta\phi=0$ and the double hump at $\Delta\phi=\pi$ are seen along the $\Delta\eta$ axis. The narrow peak at $(\Delta\eta,\Delta\phi)\sim(0,0)$ is due to jets and other short range correlations and is removed by applying a $|\Delta\eta|>2$ cut. Panel b) shows the one-dimensional correlation function in $\Delta\phi$ which only contains contribution from the long range structures (in $\Delta\eta$). The ridge and the double hump are clearly visible in the 1-D correlation. As the trigger and partner $\pT$ ranges are chosen to be the same, Eq. \ref{eq_scaling_relation} reduces to:

\begin{equation}
v_{n,n}(\pT^\mathrm{a},\pT^\mathrm{a})=v_{n}(\pT^\mathrm{a})^{2}
\end{equation}

\noindent which is used to obtain the $v_{n}(\pT)$. In Fig. \ref{fig_2pc_method}, the $v_{n}$ are plotted up to $n=15$, but the analyis is limited to $n=6$ as for higher $n$ the systematic and statistical errors are too large for the $v_{n}$ values to have any significance.

\begin{figure}
\centering
\includegraphics[width=0.8\paperwidth]{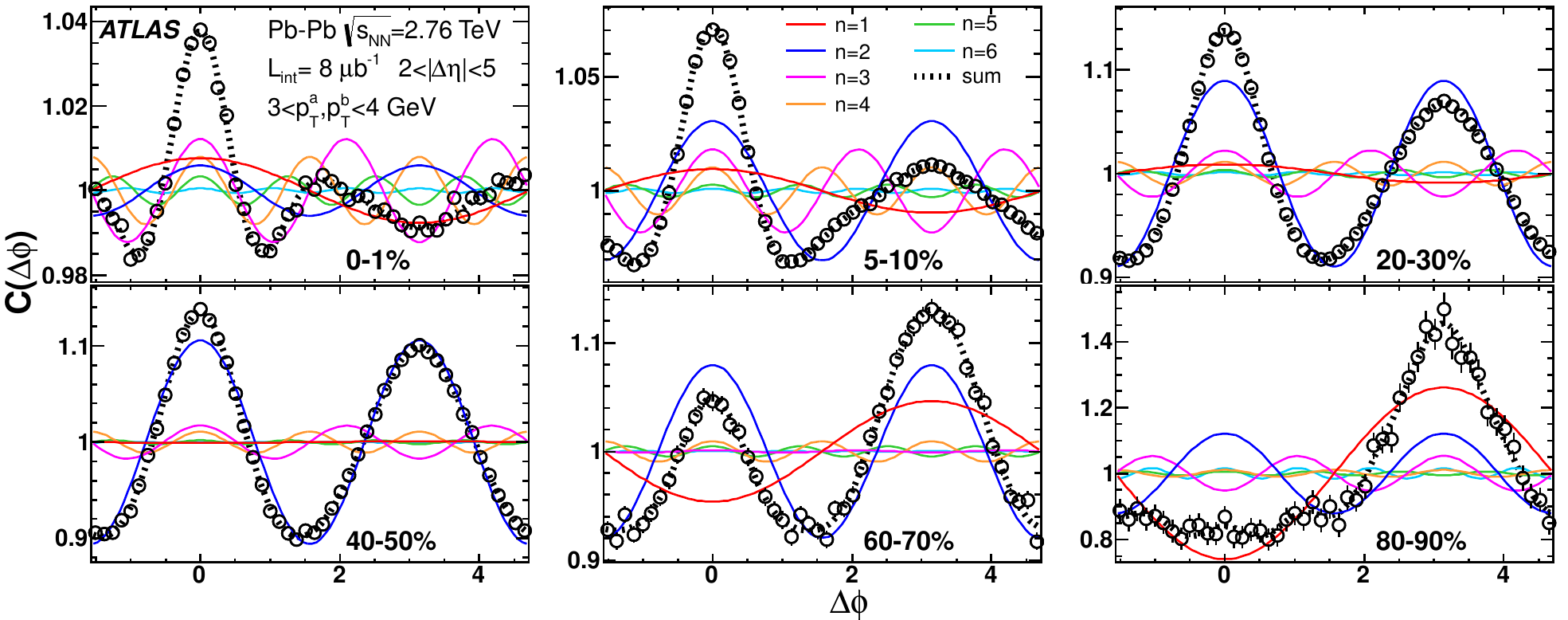}
\caption{Centrality dependence of $\Delta\phi$ correlations for $\pT^\mathrm{a},\pT^\mathrm{b}\in(3,4)$GeV and $\Delta\eta\in(2,5)$. The superimposed solid lines (thick-dashed lines) indicate contributions from individual vn,n components (sum of the first six components).}
\label{fig_2pc_cent_dep}
\end{figure}

The 2PC results can be used to check where the factorization (Eq. \ref{eq_scaling_relation}) breaks. If collective flow dominates the 2PC, then the near-side peak must be larger than the away-side peak. From Fig. \ref{fig_2pc_cent_dep} it is seen that this is roughly true up to $50\%$ centrality (for $\pT^\mathrm{a}$,$\pT^\mathrm{b}\in(3,4)$ GeV) beyond which the away-side peak becomes larger indicating a break-down of the $v_{n,n}$ factorization. 

\begin{figure}
\centering
\includegraphics[width=0.6\paperwidth]{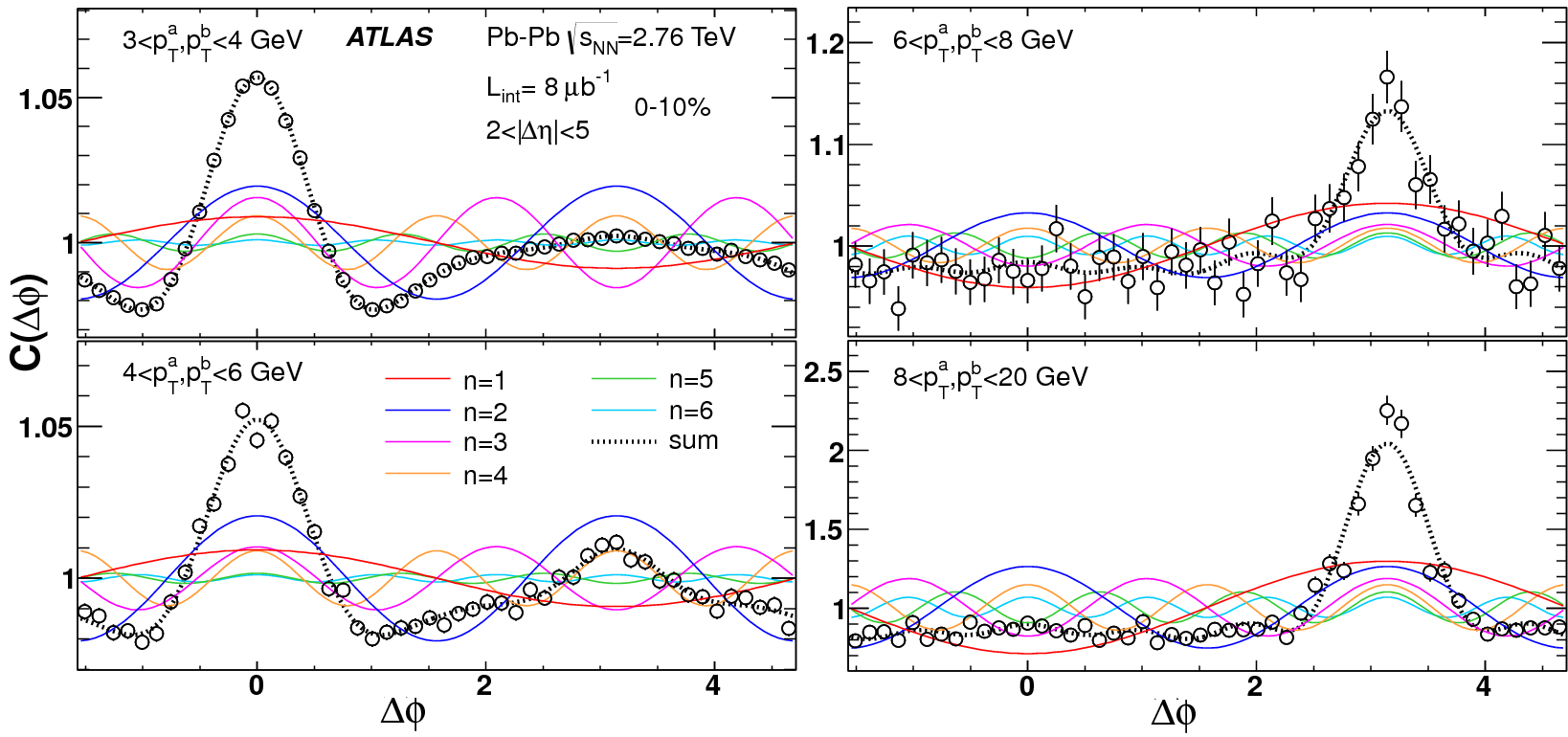}
\caption{Fixed-$\pT$ correlation functions in the (0-10)\% centrality interval for several $\pT$ ranges with $2<|\Delta\eta|<5$.}
\label{fig_2pc_pt_dep}
\end{figure}

In Fig. \ref{fig_2pc_pt_dep}, the $\pT$ evolution of the correlations is shown for $(0-10)\%$ central collisions. At low $\pT$ the correlation is driven by flow and the near-side peak is larger than the away-side. However, for $\pT^\mathrm{a}$,$\pT^\mathrm{b}>6$ GeV, the 2PCs are dominated by the away-side jet peak. Thus it is clear that at high $\pT$ ($>6$ GeV) and in peripheral collisions ($>50\%$ centrality), non-flow effects become important and Eq. \ref{eq_scaling_relation} is not expected to hold.

\section{Comparison of $v_{n}$ obtained from 2PC and EP methods}

\begin{figure}
\centering
\includegraphics[width=0.6\paperwidth]{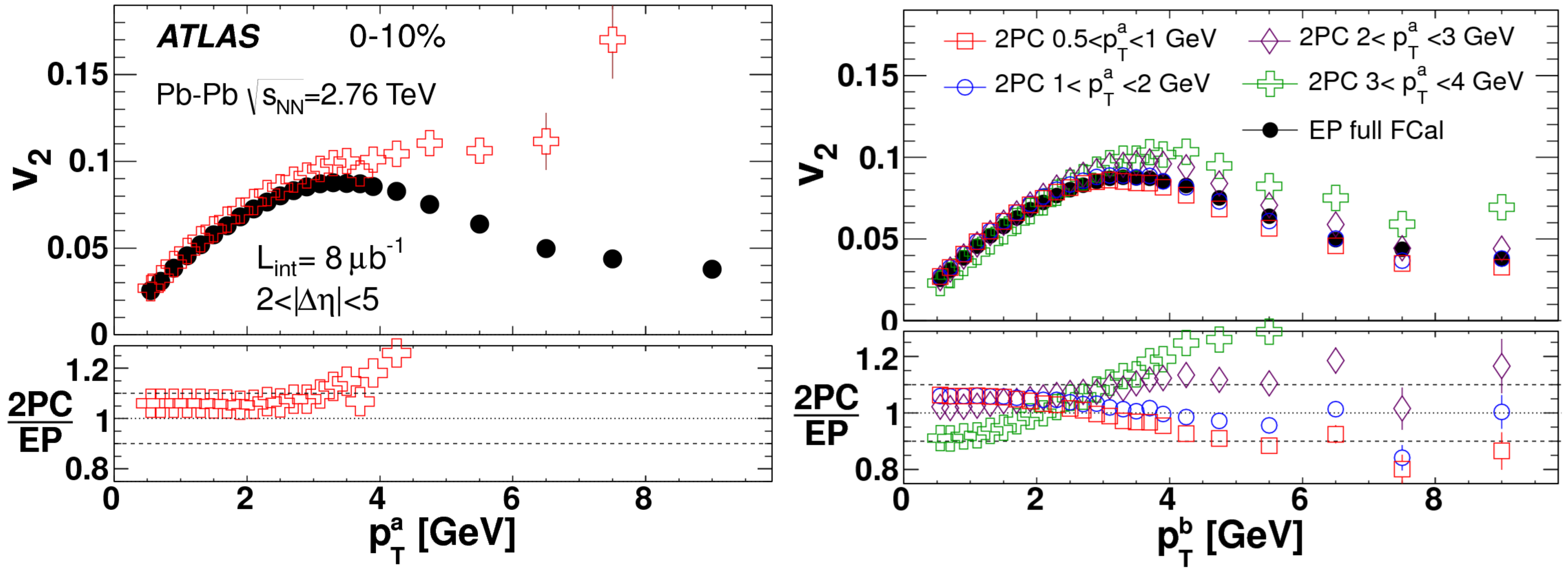}

\caption{Left panels: comparision between the fixed-$\pT$ $v_2(\pT)$ (or $v_3(\pT)$) and EP $v_2(\pT)$ (or $v_3(\pT)$. Right panels: Comparison of $v_2(\pT^\mathrm{b})$ ($v_3(\pT)$) obtained for four different reference $\pT^\mathrm{a}$ ranges (0.5-1, 1-2, 2-3, 3-4 GeV) with the EP values. The error bars indicate the statistical uncertainties only. The dashed lines in the ratio plots indicate a $\pm 10\%$ band to guide eye. All results are for the (0-10)\% centrality interval.}
\label{fig_compare_ptdep}
\end{figure}

The left panels of Fig. \ref{fig_compare_ptdep}, compare the $v_{2}$ from fixed-$\pT$ 2PC method with those from the EP method for $(0-10)\%$ central collisions. The two methods agree within $5\%$ to $15\%$ for $v_{2}$ for $\pT<4$ GeV. Deviations are observed for $\pT>4$ GeV, presumably due to contributions from the away-side jet.

The $v_n$ values obtained from fixed-$\pT$ correlations can be cross-checked using correlations where the trigger and partner have different $\pT^\mathrm{a}$ and $\pT^\mathrm{b}$ values. Such correlations are termed as mixed-$\pT$ correlations. The $v_{n}(\pT^\mathrm{b})$ of the partner can be obtained as:

\begin{equation}
v_{n}(\pT^\mathrm{b})=\frac{v_{n,n}(\pT^\mathrm{a},\pT^\mathrm{b})}{v_{n}(\pT^\mathrm{a})}
\label{eq_mixed_relation}
\end{equation}

\noindent Equation  \ref{eq_mixed_relation} can be checked by measuring the same $v_{n}(\pT^\mathrm{b})$ for different $v_{n}(\pT^\mathrm{a})$. This is illustrated for $v_{2}$ in the right panel of Fig. \ref{fig_compare_ptdep}. It is seen that factorization of $v_{n,n}$ works well for $n=2$ and the values of $v_{2}(\pT^\mathrm{b})$ are reasonably independent of $\pT^\mathrm{a}$. Further it is seen that when $\pT^\mathrm{a}$ is below 3 GeV, the agreement of the 2PC values with the EP method extend out to much higher $\pT^\mathrm{b}$. This shows that as long as the reference $\pT^\mathrm{a}$ is low, the factorization relation is valid and hence the $v_{n}$ can be measured to high $\pT$ via the 2PC method. This factorization between hard and soft particles is expected since the high $\pT$  particles, due to path-length dependent energy loss, are correlated with the same geometry that drives the collective expansion at low $\pT$  \cite{bib_jjia_paper}.

\begin{figure}
\includegraphics[width=0.59\paperwidth]{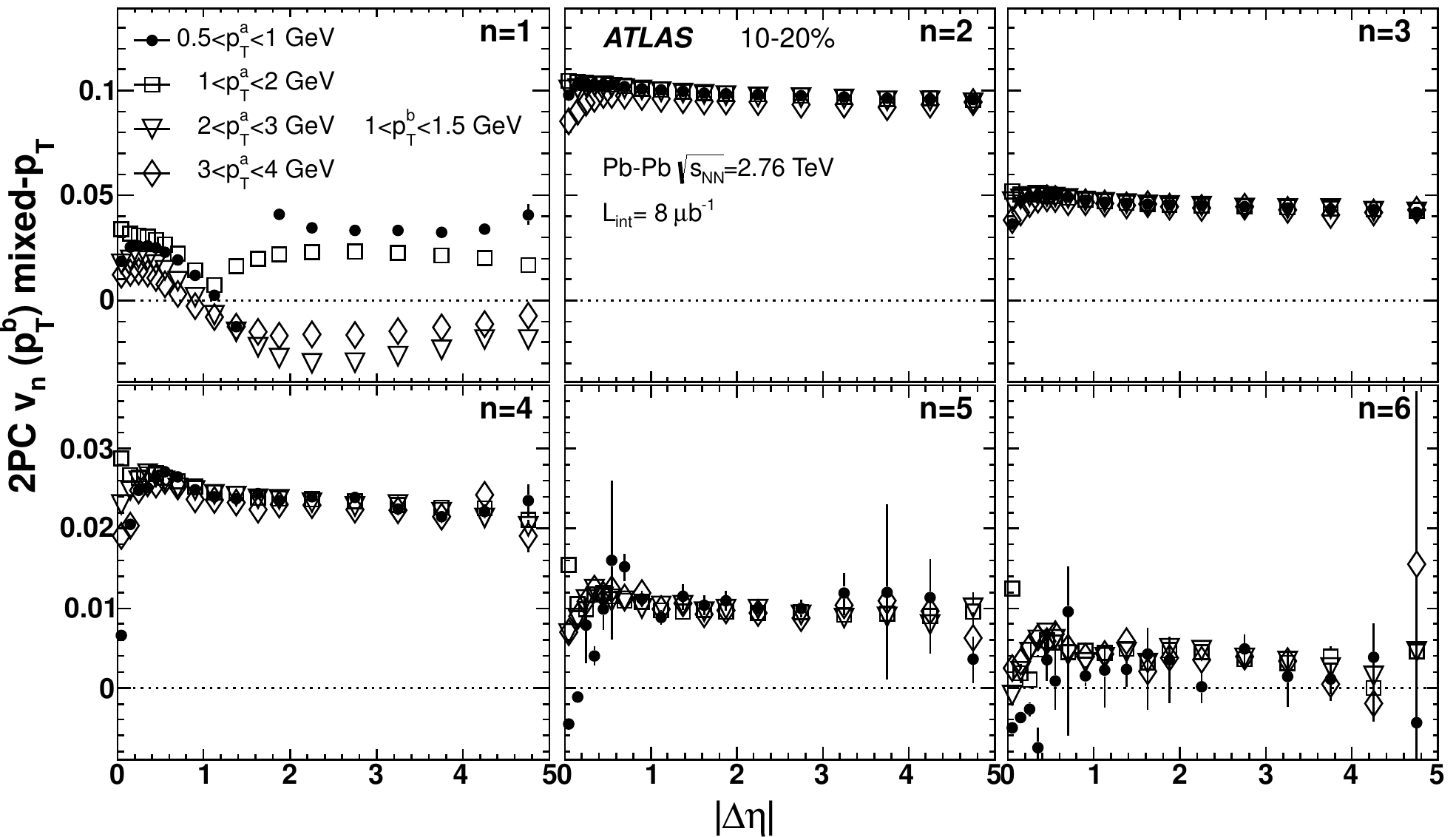}
\begin{minipage}[b]{9pc}
\caption{\label{fig_deta_dep}$v_n(\pT^\mathrm{b})$ vs. $|\Delta\eta|$ for $\pT^\mathrm{b}\in(1,1.5)$ GeV, calculated for four different reference $\pT^\mathrm{a}$ ranges (0.5-1, 1-2, 2-3, and 3-4 GeV). The error bars indicate the statistical uncertainties only.\vspace{140pt}}
\end{minipage}
\end{figure}

In Fig. \ref{fig_deta_dep} the validity of the factorization is checked for all harmonics as a function of $|\Delta\eta|$ for four different $\pT^\mathrm{a},\pT^\mathrm{b}$ combinations. For harmonics $v_2$ to $v_6$ and $|\Delta\eta|>1.0$, all trigger-partner combinations give nearly identical values of $v_n(\pT^\mathrm{b})$ even though the trigger $v_n$ values vary over a large range. This validates the factorization for $n=2-6$.

\section{Extracting dipolar $v_1$ from $v_{1,1}$}

\begin{figure}[h]
\includegraphics[width=0.8\paperwidth]{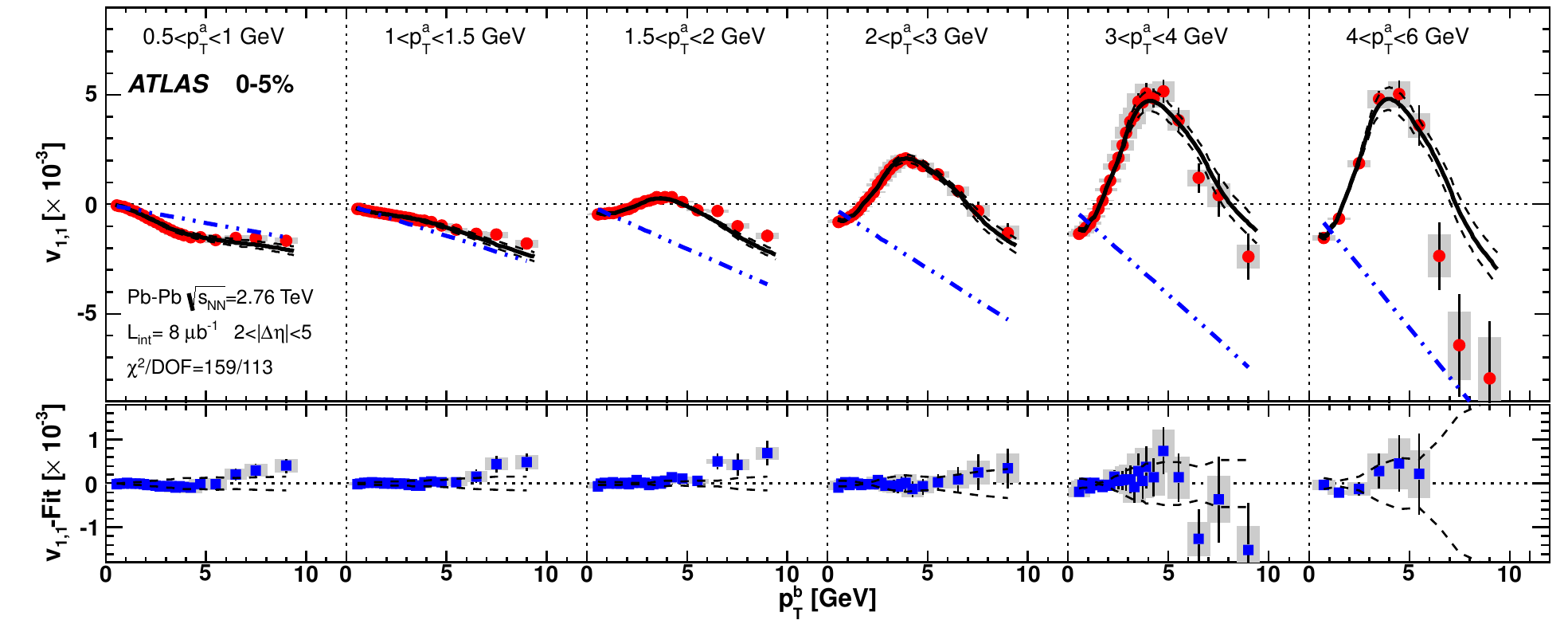}
\caption{Global fit to the $v_{1,1}$ data for the (0-5)\% centrality interval. The fit is performed simultaneously over all $v_{1,1}$ data points in a given centrality interval, which are organized as a function of $\pT^\mathrm{b}$ for various $\pT^\mathrm{a}$ ranges (indicated at the top of each panel), with shaded bars indicating the correlated systematic uncertainties. The fit function and the associated systematic uncertainties are indicated by the thick-solid lines and surrounding dashed lines, respectively. The dot-dashed lines intercepting each dataset (with negative slope) indicate estimated contributions from the momentum conservation component. The lower part of each panel shows the difference between the data and fit (solid points), as well as the systematic uncertainties of the fit (dashed lines)}
\label{fig_v11_fit}
\end{figure}

The first panel of Fig. \ref{fig_deta_dep} shows that for $v_{1,1}$ the factorization does not hold for any value of the $\Delta\eta$ gap. This is because the $v_{1,1}$ is strongly influenced by global momentum conservation (GMC), which modifies the factorization for $v_{1,1}$ to \cite{bib_v1_momentum_conservation}:

\begin{equation}
v_{1,1}(\pT^\mathrm{a},\pT^\mathrm{b},\eta^\mathrm{a},\eta^\mathrm{b})=v_{1}(\pT^\mathrm{a},\eta^\mathrm{a})v_{1}(\pT^\mathrm{a},\eta^\mathrm{a})-\frac{\pT^\mathrm{a}\pT^\mathrm{b}}{M\langle \pT^{2}\rangle}
\label{eq_v11_full}
\end{equation}

\noindent where, $M$ is the multiplicity in the event. The GMC term in the above equation is the leading order approximation for momentum conservation and is important at high $\pT$ and when the multiplicity is low. The $v_{1}(\pT^\mathrm{a},\eta^\mathrm{a})$ can be decomposed into rapidity-even and rapidity-odd components. The rapidity-odd component of $v_{1}$ is due to sideward deflection of the colliding nuclei and is small at mid-rapidity (less than 0.005 for $|\eta|<2$), hence its contribution to $v_{1,1}$ is small ( $<2.5\times10^{-5}$ ). The rapidity-even component comes from a dipole asymmetry due to fluctuations in in the initial geometry and like the other $v_{n}$ is expected to be large and have a weak $\eta$ dependence \cite{bib_dipole_paper}. In this case, Eq. \ref{eq_v11_full} simplifies to:

\begin{equation}
v_{1,1}(\pT^\mathrm{a},\pT^\mathrm{b})=v_{1}(\pT^\mathrm{a})v_{1}(\pT^\mathrm{a})-c\pT^\mathrm{a}\pT^\mathrm{b}
\end{equation}

\noindent where $c=\frac{1}{M\langle \pT^{2}\rangle}$. The $v_{1}(\pT)$ are obtained as fit parameters by fitting $v_{1,1}(\pT^\mathrm{a},\pT^\mathrm{b})$ according to the above equation with $c$ as an additional fit parameter. Figure  \ref{fig_v11_fit} shows the results of the fit for (0-5)\% central collisions. The lower panels in Fig. \ref{fig_v11_fit} show the difference between the fit and the data indicating that the two component fit works very well.

\begin{figure}[h]
\includegraphics[width=0.53\paperwidth]{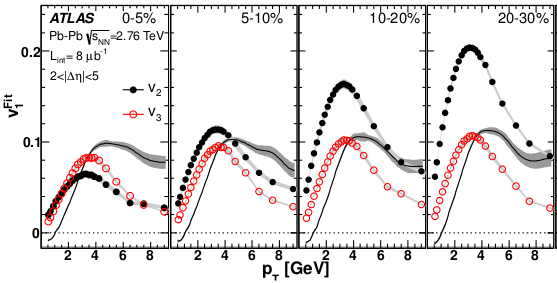}
\begin{minipage}[b]{12pc}
\caption{\label{fig_v1_values}The line shows $v_1^{Fit}$ as a function of $\pT$ for various centrality intervals. Also shown are the $v_2$ and $v_3$ values for comparison. The shaded bands indicate the total uncertainty in $v_1^{Fit}$. \vspace{110pt}}
\end{minipage}
\end{figure}

Figure  \ref{fig_v1_values} shows the dipolar $v_1(\pT)$ values obtained from the two component fit with the $v_2$ and $v_3$ values. A large dipolar $v_1$ is observed and is comparable to $v_3$, indicating a significant dipolar asymmetry in the initial geometry. The $\pT$ dependence of the dipolar $v_1$ is similar to other harmonics: it increases up to intermediate $\pT$ then decreases, which is expected as its origin is the same as the other harmonics. The negative dipolar $v_1$ seen at low $\pT$ is expected from hydro calculations \cite{bib_dipole_paper}.

\begin{figure}
\includegraphics[width=0.5\paperwidth]{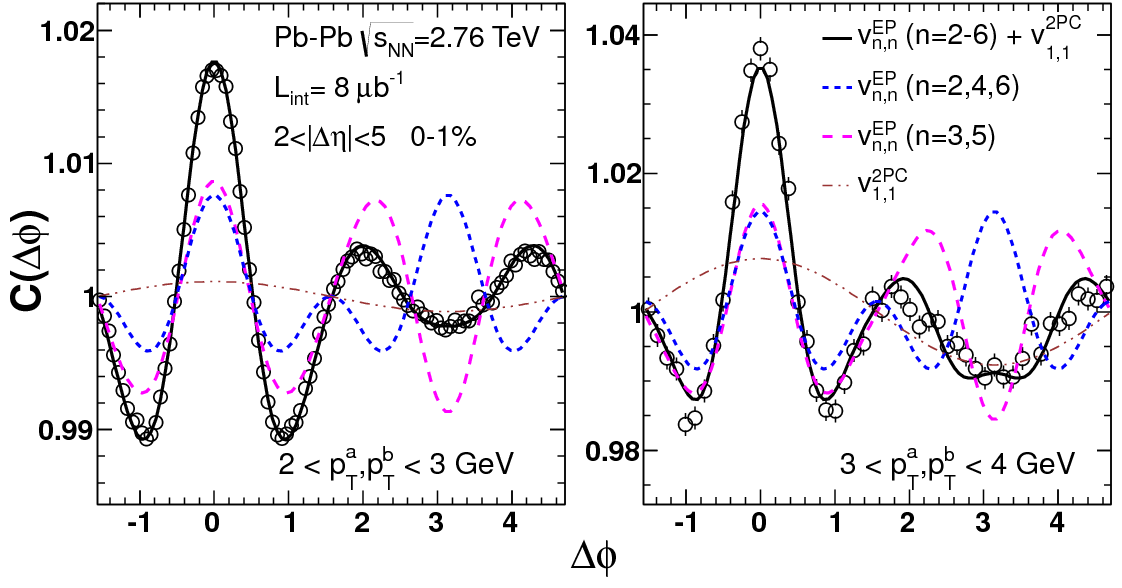}
\begin{minipage}[b]{14pc}
\caption{\label{fig_recover_cone} Measured correlation functions compared with those reconstructed from $v_2$-$v_6$ measured with the EP method and $v_{1,1}$ with the 2PC method in (0-1)\% centrality interval for two $\pT$ ranges. The contributions from n=1, n=3,5 and n=2,4,6 are shown separately. The error bars indicate the statistical uncertainties.\vspace{90pt}}
\end{minipage}
\end{figure}

  The agreement between the 2PC and EP methods implies that the structures of the two-particle correlation at low $\pT$ and large $\Delta\eta$ mainly reflect collective flow. This is verified explicitly by reconstructing the correlation function as:
  
\begin{equation}
C^{\mathrm{reco}}(\Delta\phi)=N_0^{\mathrm{2PC}} \biggl( 1+2v^{\mathrm{2PC}}_{1,1}cos(\Delta\phi)+ 2\sum_{n=2}^{6}v_n^{\mathrm{EP}}v_n^{\mathrm{EP}}cos(n\Delta\phi) \biggr)
\label{eq_reco_corr}
\end{equation}

\noindent where $N_0^{\mathrm{2PC}}$ and $v^{\mathrm{2PC}}_{1,1}$ are the average of the correlation function and first harmonic coefficient from the 2PC analysis, and the remaining coefficients are calculated from the $v_n$ measured from the event plane method. Figure  \ref{fig_recover_cone} compares two measured 2PCs to the corresponding reconstructed correlations (Eq. \ref{eq_reco_corr}) showing excellent agreement between the two. The 2PC in the right panel of Fig. \ref{fig_recover_cone} has a large $v_{1,1}$ component which contributes to the double hump, however Fig. \ref{fig_v1_values} shows that a large fraction of the $v_{1,1}$ at this $\pT$ comes from the dipolar $v_1$. This demonstrates that the ridge and cone are a manifestation of single particle $v_1$-$v_6$.

\section{Event plane correlations}

Further insight into initial geometry can be obtained by studying correlations between the $\Phi_n$. The correlations between two angles $\Phi_n$ and $\Phi_m$ are described by the differential distribution $\frac{dN_{\mathrm{events}}}{dk(\Phi_n-\Phi_m)}$, where $k$ is the lowest common multiple of $n$ and $m$. This distribution can be expanded as a Fourier series as \cite{our_ep_paper,our_ep_paper2}:
\begin{equation}
\frac{dN_{\mathrm{events}}}{dk(\Phi_n-\Phi_m)}=1+2\Sigma_{j=1}^{\infty} V_{n,m}^j\cos(jk(\Phi_n-\Phi_m))
\end{equation}

\noindent where the Fourier coefficients $V_{n,m}^j$ quantify the strength of the correlations. The measured Fourier coefficients need to be corrected to account for the detector resolution effects \cite{bib_EP_note}. 

\begin{figure}
\includegraphics[width=0.8\paperwidth]{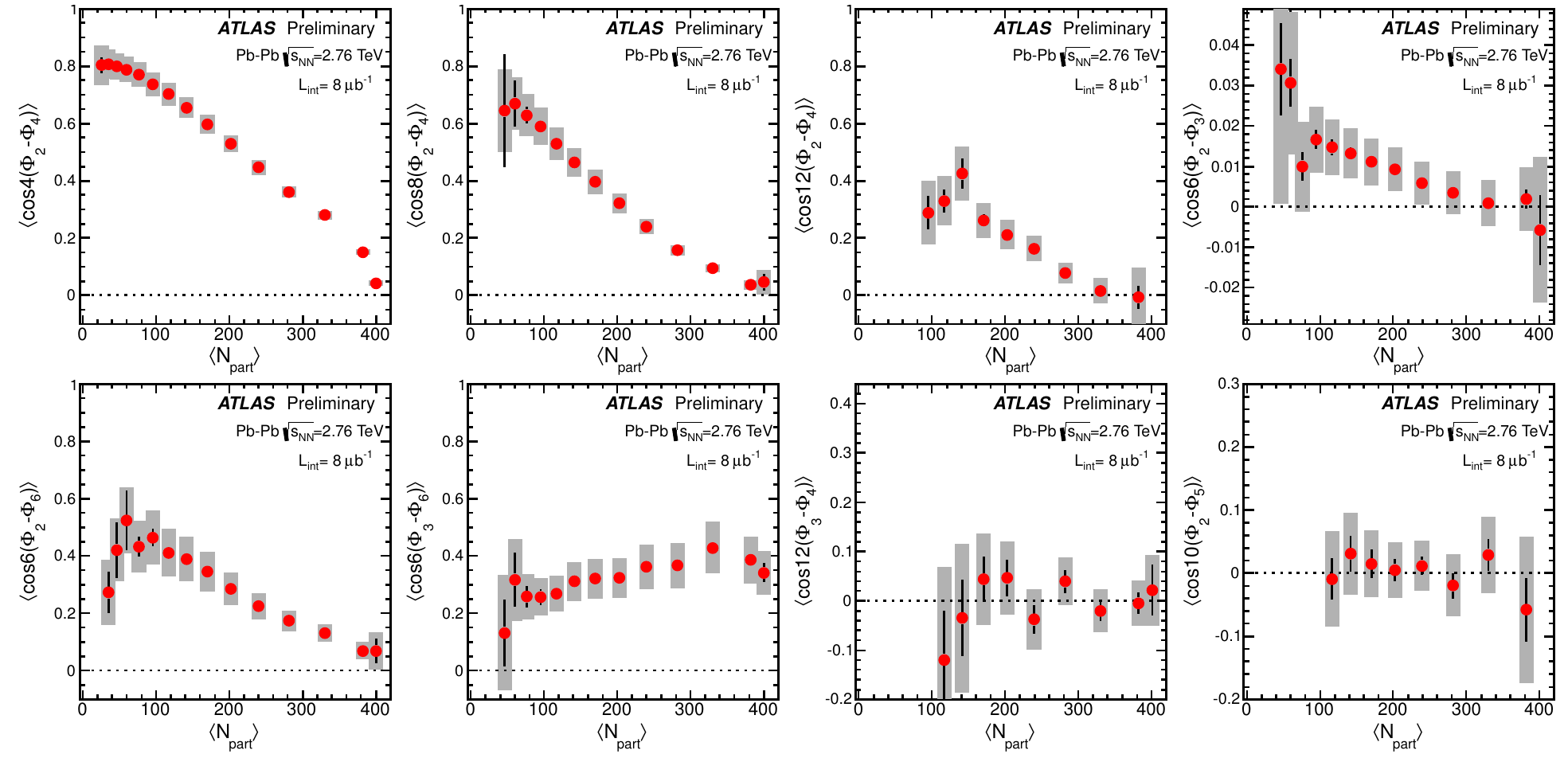}
\caption{Two-plane correlators $\langle\cos jk(\Phi_n-\Phi_m\rangle)$ as a function of $\langle N_{\mathrm{part}}\rangle$. The middle two panels in the top row have $j=2$ and $j=3$, respectively, while all other panels have $j=1$. The error bars and shaded bands indicate the statistical and systematic uncertainties respectively.}
\label{fig_two_plane_corr}
\end{figure}

The two-plane correlations are summarized in Fig. \ref{fig_two_plane_corr}. The first three panels show the $j=1,2$ and 3 moments of the  $4(\Phi_2-\Phi_4)$ correlation as a function of the number of participating nuclueons $\langle N_{\mathrm{part}} \rangle$. The correlations are small in most central collisions and increase almost linearly with decreasing $\langle N_{\mathrm{part}} \rangle$ becoming fairly large in mid-central and peripheral collisions showing a strong correlation between $\Phi_2$ and $\Phi_4$.

The fourth panel shows the $j=1$ moment of the correlation between $\Phi_2$ and $\Phi_3$. The correlations are very weak, consistent with zero in the most central collisions and increase to roughly $2\%$ in peripheral collisions. The next two panels show the $j=1$ moments of the $\Phi_2-\Phi_6$ and $\Phi_3-\Phi_6$ correlations. The $\Phi_2$ and $\Phi_3$ are weakly correlated with each other, but they individually are strongly correlated with $\Phi_6$. Also the two correlations show completely different centrality dependance : $\Phi_2-\Phi_6$ correlation increases almont linearly with decreasing $\langle N_{\mathrm{part}} \rangle$ ($i.e.$ from central to peripheral collisions) while $\Phi_3-\Phi_6$ gradually decreases. The last two panels of Fig. \ref{fig_two_plane_corr} show the $\Phi_3-\Phi_4$ and $\Phi_2-\Phi_5$ correlations respectively, which are found to be weak (less than a few \%) and consistent with zero. It is possible to generalize the two-plane correlations to correlations between three or more planes. A detailed analysis of these event plane correlations is presented in a separate talk in this conference by J. Jia \cite{bib_jjia_proceedings}.

\section{Summary}

Measurements of the flow harmonics $v_1$-$v_6$ over a large $\pT$, $\eta$ and centrality range are presented. The 2PC $v_{n,n}$ are shown to factorize into products of single particle $v_n$ for $n=2-6$ in central and mid-central collisions as long as one particle has low $\pT$ ($<3$GeV). The factorization breaks for $v_{1,1}$ due to the contribution from momentum conservation effects. Dipolar $v_1$ is extracted from the $v_{1,1}$ data via a two component fit. The extracted dipolar $v_1$ is comparable to $v_3$, indicating significant dipole asymmetry in the initial geometry. It is shown that the features in two-particle correlations for $|\Delta\eta|>2$ at low and intermediate $\pT$ ($<3$ GeV) such as the ridge and double-hump are accounted for by the $v_1$-$v_6$. 

The $v_n$ can be thought of diagonal components of a “Flow Matrix”. Studying the two and three plane correlations gives access to the off diagonal entries and beyond. A detailed set of event plane correlation measurements are presented.

This work is in part supported by NSF under award number PHY-1019387.





\bibliographystyle{elsarticle-num}
\bibliography{<your-bib-database>}

\begin{thebibliography}{00}
\bibitem{reaction_plane_defns}A. M. Poskanzer and S. A. Voloshin, Phys. Rev. C58,1671 (1998).

\bibitem{bib_visc_calc1} Z. Qiu and U. W. Heinz, Phys. Rev. C 84, 024911 (2011). 

\bibitem{bib_visc_calc2} B. Schenke, S. Jeon, C. Gale, Phys. Lett. B702, 59 (2011).


\bibitem{bib_ridge_cone1} STAR Collaboration, B. I. Abelev \textit{et al}. Phys. Rev. C 80, 064912 (2009).

\bibitem{bib_ridge_cone2} PHENIX Collaboration, A. Adare \textit{et al}., Phys. Rev. C 78, 014901 (2008).

\bibitem{bib_v3cone_} B. Alver and G. Roland, Phys. Rev. C 81, 054905 (2010) [Erratum Phys. Rev. C 82, 039903 (2010)]

\bibitem{our_ep_paper} J. Jia and S. Mohapatra, A method for studying initial geometry fluctuations via event plane correlations in heavy ion collisions, arXiv:1203.5095 [nucl-th]. 
\bibitem{our_ep_paper2} J. Jia and D. Teaney, Study on initial geometry fluctuations via participant plane correlations in heavy ion collisions: part II, arXiv:1205.3585v1 [nucl-ex].


\bibitem{atlas_detector} ATLAS Collaboration, JINST 3, S08003 (2008).

\bibitem{bib_vn_note} ATLAS Collaboration, Phys. Rev. C 86, 014907 (2012).

\bibitem{bib_EP_note} ATLAS Collaboration, ATLAS-CONF-2012-049, Measurement of reaction plane correlations in Pb-Pb collisions at $\sqrt{s_{NN}}$=2.76 TeV, http://cdsweb.cern.ch/record/1451882

\bibitem{bib_jjia_paper} J. Jia, Azimuthal anisotropy in a jet absorption model with fluctuating initial geometry in heavy ion collisions, arXiv:1203.3265v2 [nucl-th].

\bibitem{bib_v1_momentum_conservation} M. Luzum  and J. Ollitrault, Phys. Rev. Lett. 106, 102301 (2011). 

\bibitem{bib_dipole_paper} D. Teaney and L. Yan, Phys. Rev. C 83, 064904 (2011). 

\bibitem{bib_jjia_proceedings} Jiangyong Jia for the ATLAS Collaboration, this proceeding, arXiv:1208.1427v1 [nucl-ex].
 

 \end{thebibliography}



\end{document}